\documentclass[aps,prl,twocolumn,groupedaddress]{revtex4}
\usepackage{graphicx}
\usepackage{amsmath}
\usepackage{amsfonts}
\usepackage{amssymb}
\usepackage{epstopdf}

\graphicspath{{./Figures/}} 
\begin{document}


\title{A Simple Model for Identifying Critical Regions in Atrial Fibrillation}


\author{Kim Christensen$^{1}$, Kishan A. Manani$^{1,2}$, Nicholas S. Peters$^{2}$}
\affiliation{$^{1}$The Blackett Laboratory, Imperial College London, London SW7 2BW, United Kingdom \\ 
$^{2}$National Heart and Lung Institute, Imperial College London, London W12 0NN, United Kingdom}
\newcommand{\PCoupling}{\nu}
\newcommand{\PDefect}{\delta}
\newcommand{\Pcoupling}{\nu}
\newcommand{\Pdefect}{\delta}
\newcommand{\Dysf}{\epsilon}
\newcommand{\RP}{\tau}
\newcommand{\lm}{\ell_{\text{min}}}
\newcommand{\pv}{p_{\nu}}
\newcommand{\dxprime}{\Delta x^{\prime}} 
\newcommand{\dtprime}{\Delta t^{\prime}}
\newcommand{\dyprime}{\Delta y^{\prime}}
\newcommand{\RPprime}{\tau^{\prime}}
\newcommand{\CVxprime}{{\theta_{x}}^{\prime}}

\date{\today}

\begin{abstract}

Atrial fibrillation (AF) is the most common abnormal heart rhythm and the single biggest cause of stroke.
Ablation, destroying regions of the atria, is applied largely empirically and can be curative but with
a disappointing clinical success rate. We design a simple model of activation wavefront propagation on an anisotropic structure mimicking the branching network of heart muscle cells.  This integration of phenomenological dynamics and pertinent structure shows how AF emerges spontaneously when the transverse cell-to-cell coupling decreases, as occurs with age, beyond a threshold value. We identify critical regions responsible for the initiation and maintenance of AF, the ablation of which terminates AF. The simplicity of the model allows us to calculate analytically the risk of arrhythmia and express the threshold value of transversal cell-to-cell coupling as a function of the model parameters. This threshold value decreases with increasing refractory period by reducing the number of critical regions which can initiate and sustain micro-reentrant circuits. These biologically testable predictions might inform ablation therapies and arrhythmic risk assessment. 
\end{abstract}


\maketitle



Atrial fibrillation (AF) is a poorly understood arrhythmia where multiple activation wavefronts are observed to propagate continuously in the atrial muscle tissue in an apparently random manner \cite{Konings1994}. The activation wavefronts meander, divide and collide with mutual extinction and re-initiation. This manifests in patients as episodes of variable duration, from short self-terminating episodes (paroxysmal AF), which typically get longer with time to become self-sustaining (persistent AF) \cite{Kerr2005}.  The incidence of AF increases with age and is strongly associated with the accumulation of fibrosis \cite{de2011fibrosis}, however, the exact mechanisms underlying AF are still not understood. 

The relationship between heart muscle architecture (structure) and the propagation of activation wavefronts, i.e., the relationship between structure and function, may be a key mechanism. Rotating spiral wavefronts of activity (rotors) have been observed in isolated cardiac tissue \cite{davidenko1992stationary} and recently in human AF \cite{Narayan2012}, suggesting critical regions for persistence of AF. But why rotors should form and their relationship to the underlying structure and its arrhythmogenic characteristics are unknown. Heterogeneity in structural and electrophysiological tissue properties (e.g. fibrosis \cite{spach2007mounting}, orifices \cite{haissaguerre1998}, action potential duration \cite{satoh1996unequal}, restitution \cite{kim2002action}, tissue thickness \cite{eckstein2013transmural}) have all been implicated in the perpetuation of AF. 

Ablating atrial tissue extensively and empirically via catheter electrodes may cure AF in some cases \cite{haissaguerre1998, Nademanee2004}, but an inability to identify the specific regions critical to the persistence of AF has resulted in failure to improve on disappointing clinical outcomes. Thus, enhancing mechanistic understanding, identification of critical causative regions and prediction of arrhythmic risk for preventative strategies requires a novel approach.

Heart muscle cells are discrete and so it is natural to use cellular automata to
model activation wavefronts as was done in some of the earliest mathematical models of AF \cite{wiener1946mathematical,moe1964computer}. These studies investigated the effect of heterogeneity in functional parameters (e.g., refractory period) and macroscopic obstacles on the dynamics of wavefronts. In agreement with physiological experiments, the studies showed that an increase in hetereogeneity in refractory period resulted in greater wavefront complexity. Bub, Shrier and Glass devised cellular automata models of cardiac conduction and experimentally
compared the behaviour of the model to cardiac cell monolayers \cite{bub2002spiral, bub2005global}.
Bub {\it et al.} have also investigated the relationship between structural heterogeneity and the
stability of activation wavefronts in isotropically coupled tissue for comparison with cardiac monolayers \cite{Bub2002}.

In contrast, we have investigated the effect of anistropic structural heterogeneity in the underlying lattice of cells to  mimic the lateral uncoupling of myocardial strands by fibrosis in cardiac muscle \cite{de2011fibrosis}. The level of structural heterogeneity is controlled by $\Pcoupling$,  a model parameter that determines the fraction of cells with lateral cell-to-cell couplings. Qualitatively, the model reproduces many of the known characteristics of real AF. For example, we find that with low levels of structural heterogeneity wave propagation remains planar. As the level of structural heterogeneity is increased beyond a threshold value, we observe spontaneous localised disruption of propagation and progressive degeneration to fibrillation, either self terminating or sustained. Hence, the simple phenomenological model reproduces the observation that age related changes, that is, an increase in structural anisotropy \cite{spach1986relating,de2011fibrosis}, might spontaneously induce and sustain AF. We stress that AF behaviour is initiated spontaneously in this model as a result of the pertinent underlying structure of the tissue as opposed to a combination of rapidly paced tissue and functional heterogeneity or macroscopic obstacles.

The simplicity of the model provides mechanistic insight into the initiation and maintenance of AF. We can characterise and identify the composition of local regions critical to the initiation and maintenance of AF and show that ablation of such critical regions may terminate AF. Furthermore, we predict the existence of a critical level of heterogeneity beyond which AF occurs. Due to the simplicity of the model, we can analytically calculate the risk of AF as a function of model parameters. This analytic expression reveals that the threshold value of structural anisotropy decreases with increasing refractoy period. Bringing the threshold $\Pcoupling^{\star}$ below $\Pcoupling$ by increasing the refractory period prevents micro-reentry circuits from persisting or forming because the wavefront in the circuit collides with its unexcitable tail.

Our model of AF is complementary to biophysical models where subcellular processes,
such as ion channel activity, are taken into account \cite{Clayton2011,Clayton2001}. Biophysical models of cardiac tissue commonly consist of a system of partial differential equations (PDEs), known as the bidomain model, coupled nonlinearly to a system of ordinary differential equations modelling the cell membrane dynamics. With simplification such biophysical models are amenable to analytic calculations \cite{Sundnes2006}. E.g., Keener \cite{keener1986geometrical,keener1988dynamics} and Biktashev \cite{biktasheva2009computation} have derived analytic calculations for spiral wave dynamics on a reduced form of these equations. PDE based models assume a continuous medium while our main investigation is on the effects of anisotropic structural heterogeneity for which the discreteness of the tissue may be important \cite{Spach1994, spach1997microfibrosis, spach1996stochastic}. 

A healthy atrial area is $ {L^{'}} \times {L^{'}} \approx 20cm^{2}$ \cite{Lang2005, maceira2010reference}. Atrial muscle tissue consists of nearly cylindrical myocytes of length $\dxprime \approx 100 \mu$m and diameter $\dyprime \approx 20 \mu$m packed in an irregular brick-wall like pattern \cite{Luke1991, Verheule2003}. Cells are coupled primarily from end-to-end along the longitudinal axis of the cell with little side-to-side coupling, thus forming a branching cable-like structure \cite{Luke1991, Nakamura2011}. The structure and strength of these connections can change, e.g., fibrosis can uncouple cells, primarily in a side-to-side fashion \cite{Luke1991}. To represent this branching network of interacting cells, each cell is coupled to its neighbours longitudinally and to its neighbours transversally with probability $\PCoupling$. Structurally, the substrate can be pictured as a set of aligned longitudinal cables of single cell thickness which are coupled to neighbouring cables with frequency $\PCoupling$. The structure is a 2D sheet. A 2D sheet is a good first approximation because the atrial muscle wall is relatively thin, $2.5\text{mm}$ \cite{Nakamura2011}. However, one might extend the model to 3D by coupling adjacent 2D sheets of tissue. To mimic the basic topological features of an atrium we apply periodic boundary conditions vertically and open boundary conditions horizontally to the 2D model. This provides a cylindrical topology which is a convenient substrate to study the evolution of wavefronts as our focus is on investigating the effects of increasing anisotropic structural heterogeneity, rather than the detailed effects of atrial anatomy. Once the substrate is generated, it remains fixed throughout the numerical simulation. Thus coupling related changes in the structure are incorporated into the substrate by the single variable $\PCoupling$.

\vspace*{-0.25cm}
\begin{figure}[!ht]\centering
\includegraphics[width=0.48\textwidth]{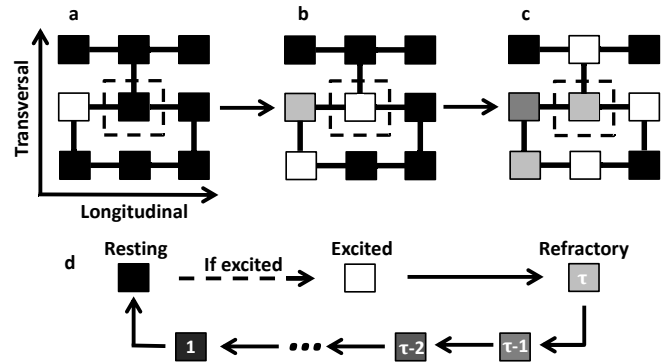}
\vspace*{-0.5cm}
\caption{ (a-c) A coupling is represented by a link between two cells. Cells are always coupled longitudinally and with frequency $\PCoupling$ transversely. (a-b)  A resting cell (black) will become excited (white) in the next time step if at least one of its neighbouring coupled cells are excited. (b-c) Once a cell is excited it will enter a refractory state (grey scale) for a duration of time $\RP$. (d) The time course of a cell once it has been excited by an excited neighbouring coupled cell}
\label{Fig:ModelDynamics}
\end{figure}

The coordinated contraction of the heart is a result of a regular propagating wavefront of activation originating at the heart's natural pacemaker. The ionic currents which determine the activation of a cell are not explicitly considered in the model. In its simplest representation, a cell may be in one of three states: resting (repolarised), excited (depolarising) or refractory, see Fig.~1. An excited cell induces neighbouring resting cells to become excited and the wavefront is a coherent propagation of this excitation throughout the tissue. The time taken for a cell to depolarise, $\dtprime \approx 0.6$ms, is much shorter than the refractory period, $\RPprime \approx 150$ms. Cells along the left boundary are pacemakers and self-excite at a fixed period $T^{\prime} = 300\text{-}1000$ms to mimic normal cardiac rhythm.

Cellular electrical dysfunction of any cause introduces a degree of noise into the excitation process. A fraction $\PDefect$ of randomly chosen cells in the substrate are assigned to be ‘dysfunctional’. Such cells have a finite probability, $\Dysf$, of not exciting in response to an excited neighbour and hence can block the transmission of excitation.

\begin{figure}[!ht]\centering
\includegraphics[width=0.48\textwidth]{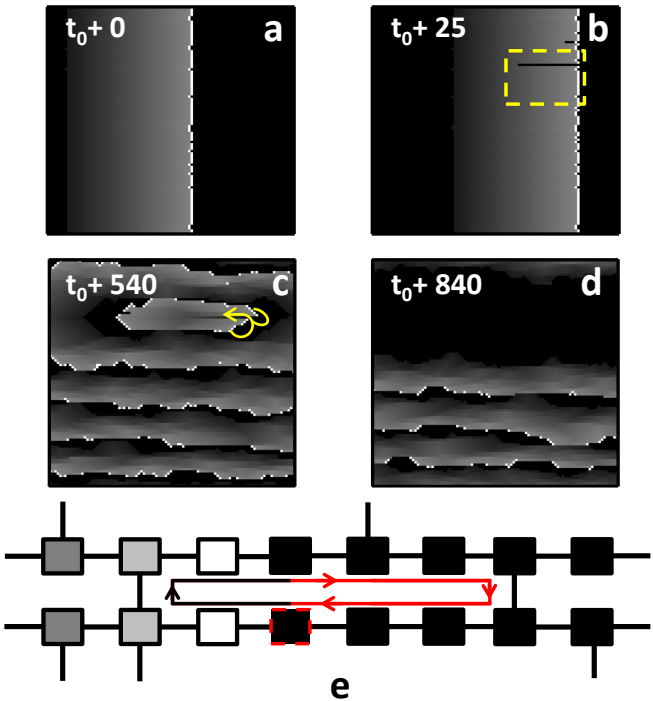}
\vspace*{-0.5cm}
\caption{Initiation and self-termination of fibrillation in a simulation with $L=200$, $\RP=50$, $\Dysf=0.05$, $\PDefect=0.05$, and $\PCoupling=0.18$. See Supp. Fig. 1 for the full image and animation. The pacemaker cells self-activate with a period of $T = 220$. (a) A planar wavefront induced by pacemaker cells along the left boundary a time $t_0$. (b) An opening occurs in the refractory wake (i.e, wave-break) and a wave of excitation leaks back on itself. (c) A reentrant circuit forms as a result of this breakthrough. (d) The reentrant circuit self-terminates and planar wave propagation resumes. (e) A critical region which can induce a rotor. The dysfunctional cell (dotted boundary) in the bottom cable may block the propagation wavefront. When the wavefront in the top cable reaches the rightmost vertical connection, it induces activity in the bottom cable that will also propagate retrograde. If the relevant path length (red line) of the circuit is greater than the refractory period, the cell to the left of the dysfunctional cell will be in a resting state when the retrograde wavefront arrives and a reentrant circuit of activation forms. The reentrant circuit terminates when the dysfunctional cell blocks the retrograde wavefront. This is the simplest example of a critical region which will induce fibrillation-like behaviour in the model.}
\label{Fig:ModelDynamics}
\end{figure}

Translating real tissue values into the model yields $L=L^{'} / \dxprime = 1000$. We coarse-grain the model by taking $\dxprime \rightarrow b\dxprime$, where $b$ is the number of cells within a unit of space in the model. As the conduction velocity $\theta_{x}' = \dxprime / \dtprime \approx 0.2$ms$^{-1}$ is fixed we also get $\dtprime \rightarrow b \dtprime$. We take $b = 5$ which gives $L=200$ and $\RP = \RP^{\prime} / (b\dtprime) = 50$. The evolution of the system varies depending on the fractions of transverse connections $\PCoupling$ and dysfunctional cells $\PDefect$ which are the only remaining parameters. Coarse-graining the transversal connections gives: $\Pcoupling \rightarrow 1-(1-\Pcoupling)^{b}$ as this is the probability that a group of $ b $ cells have at least one vertical connection between them. The parameters of the model are system size $L \!\times\! L$, refractory period $\RP$, the fraction of dysfunctional cells $\PDefect$, the probability of dysfunction $\Dysf = 0.05$, the period of pacing $T = T^{\prime}/(b\dtprime)  = 220$ using $T^{\prime} = 660$ms and the fraction of transverse connections $\PCoupling$. 
 
When $ \Pcoupling=1 $, mimicking neonatal tissue in humans and other mammals, we observe regular wavefronts for all $ \Pdefect $, consistent with the rarity of AF observed in children. We now investigate the effect of structural heterogeneity ($ \Pcoupling<1 $). For simplicity, we consider a fixed fraction of dysfunctional cells ($\PDefect=0.05$) leaving the fraction of transverse connections, $\PCoupling$, as the only control parameter. 
 
\begin{figure}[!ht]\centering
\includegraphics[width=0.45\textwidth]{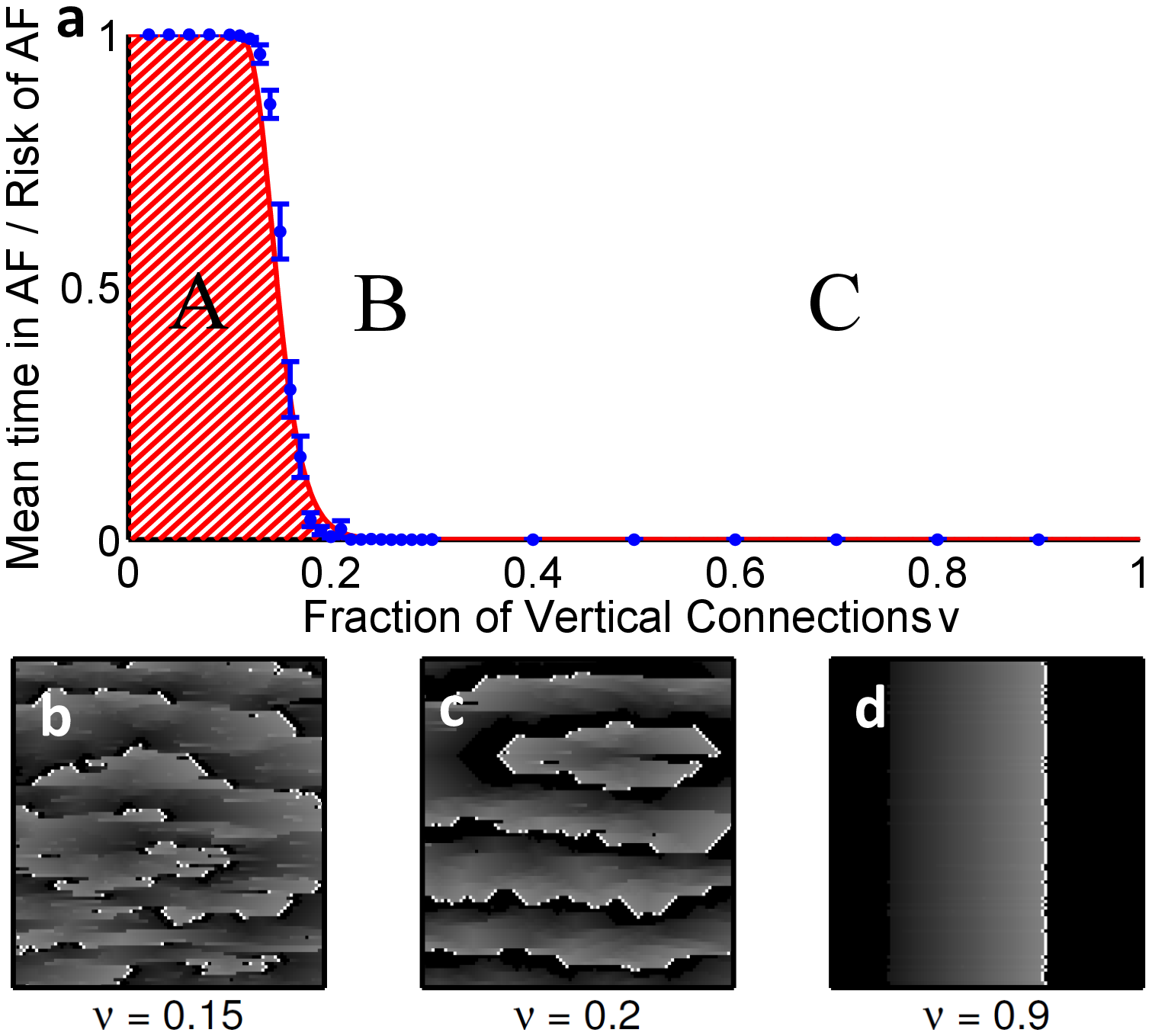}
\vspace*{-0.1cm}
\caption{(a) A phase diagram of the model with $L=200$, $\RP=50$, $\Dysf=0.05$, and $\PDefect=0.05$. Pacemaker cells self-activate with a period of $T=220$. As $\PCoupling$ decreases from 1, a transition from planar wavefronts (C) to a few self-terminating reentrant circuits (B) occurs. As $\PCoupling$ is decreased below $\Pcoupling^{\star}  \!\approx\! 1 - (\PDefect  L^{2})^{-\frac{1}{ \RP }}  \!\approx\! 0.14$ the system develops multiple self-sustaining reentrant circuits (A). The red line is the analytically calculated probability  $P_{\text{risk}} \!=\! 1-\left[1 \!-\! (1-\PCoupling)^{\RP}\right]^{\PDefect L^{2}}$ of having at least one fibrillation-inducing structure, see Fig. 2(e). The solid blue circles represent the average duration that a system displays non-planar wavefronts, averaged over 50 realisations of duration $10^6$ time steps and error bars represent one standard error of the mean. (b-d) Realisations of systems with $\PCoupling = 0.9$ - planar behaviour, $\PCoupling = 0.2$ - single self-terminating rotor and $\PCoupling = 0.15 $ - multiple self-terminating rotors. See Supp. Fig. 2 for the full image.}
\label{Fig:StoryBoard}
\end{figure}

For large fractions of transverse connections ($\PCoupling \gtrsim 0.21$) we only observe planar wavefronts. However, as $\nu$ is decreased, mimicking the onset of fibrotic interference of lateral cell-to-cell coupling, the branching nature becomes more prominent and circuits of activity may form spontaneously by wavelets of excitation leaking back through the refractory wake of the wavefront, see Fig. 2(a-c). The associated fibrillation may self-terminate, see Fig. 2(d), or become persistent. Together the structural heterogeneity and dysfunctional cells may create a reentrant circuit that constitutes a rotor, see Fig. 2(e).
\begin{figure}[!ht]\centering
\includegraphics[width=0.48\textwidth]{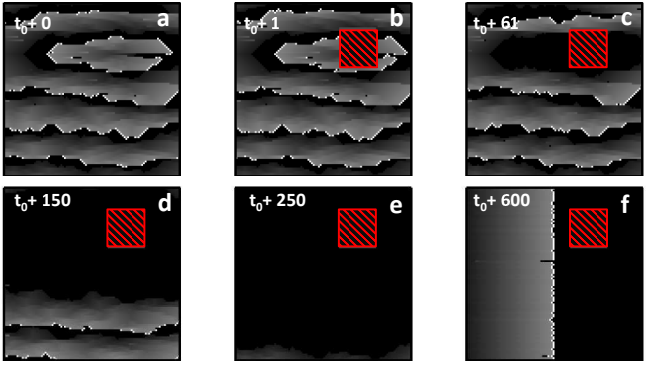}
\vspace*{-0.55cm}
\caption{(a) A rotor due to a critical structure. (b) An ablation is made by rendering a $20 \times 20$ section of tissue around the critical structure unexcitable (cross hatched red). (c-f) The ablation terminates the reentrant circuit and the system returns to a planar wavefront phase. See Supp. Fig.~3 for the full image and animation.}
\label{Fig:TimeInAF}
\end{figure}

Hence, the parameter space $\PCoupling$ may be divided into a rotor forming range and a stable range in which no rotors form, see Fig.~3. The frequency of formation and the temporal duration of rotors increases sharply from zero as the fraction of transverse connections $\PCoupling$ decreases below a threshold value $\PCoupling^{\star} \!\approx\! 0.14$. Only when $\PCoupling$ is sufficiently small will the relevant path length of the reentrant circuits exceed the refractory period, which is a necessary condition to sustain a rotor, see Fig. 2(e). Decreasing $\Pcoupling$ results in rotors forming and disassociating more rapidly resulting in a fibrillatory state, see Fig.~3. This mimics the natural onset of AF and its transition to paroxysmal and persistent atrial fibrillation. Moreover this is consistent with the increase of AF incidence with advancing age. The critical regions forming a simple reentrant circuit as displayed in Fig.~2(e) are primarily responsible for the fibrillation observed in the model. 

 In our model, we can analytically
calculate the risk of developing atrial fibrillation. We identify the risk of developing atrial fibrillation with the probability that
the $L \!\times\! L$ system has at least one portion of the reentrant circuit as displayed in red in Fig. 2(e). The probability that a cell has at least one transverse coupling is $p_{\Pcoupling} = 1 - \left(1-\Pcoupling\right)^{2}$. Given an arbitrary dysfunctional cell $i$, let $\ell_i$ denote the distance from that cell to its first neighbour to the right
that has at least one transverse coupling. 
Hence, the probability that $\ell_i < \frac{\RP}{2}$ (assuming $ \RP $ is even) 
\begin{equation}
P\left(\ell_i < \frac{\RP}{2}\right) = \sum_{\ell_i=0}^{\frac{\RP}{2}-1} (1-\pv)^{\ell_i} \pv 
			 = 1- \left(1-\Pcoupling\right)^{\RP}. 
\end{equation}
The average no. of dysfunctional cells is $\delta L^2$ and $P_{\text{risk}}$ is the complement
of all dysfunctional cells having $l_i < \frac{\RP}{2}$:
\begin{equation}
P_{\text{risk}} \!=\! 1\!-\! \left[1 \!-\! \left(1 \!-\! \Pcoupling\right)^{\RP}\right]^{\Pdefect L^{2}} 
 \!=\!  1 \!-\! [1\!-\!\left(1 \!-\! \Pcoupling\right)^{\frac{\theta_{x}^{\prime} \RP '}{b \Delta x'}}]^{ \frac{{\Pdefect L^{'}}^{2}}{b^2{\dxprime}{\dyprime}}}.
\end{equation}
\label{EQN2}
The first expression is in dimensionless units, while the second expression is in units with dimensions where $\dtprime = \dxprime / \theta_{x}^{\prime}$ where $\theta_{x}^{\prime}$ is the conduction velocity in the longitudinal direction.

 The analytically calculated risk's fit to the data, see solid red line in Fig.~3(a), confirms that the critical regions shown in Fig.~2(e) drive fibrillation in the model. By rendering the region containing this structure unexcitable, mimicking ablation, we find that the rotor terminates and planar wavefronts reappear, representing a return to regular cardiac rhythm, see Fig. 4(a-f).


The model displays a sharp transition from regular to fibrillatory behaviour as a function of the control parameter $\Pcoupling$. The existence of the transition is technically a finite-size effect because as the system size increases, the planar wavefront range (C) of the phase diagram decreases and will, eventually disappear for $ L\rightarrow \infty $. However, the threshold value $\Pcoupling^{\star}$, defined by the point of steepest slope of $P_{\text{risk}}$, increases extremely slowly with system size as $\PCoupling^* \approx 1 - (\PDefect  L^{2})^{-\frac{1}{ \RP }} \approx 1 \!-\! ( \Pdefect {L^{\prime}}^{2}/{b^2 \dxprime \dyprime})^{-b \dxprime / \CVxprime \RPprime} $, for $ \RP \gg 1$ and $ \PDefect L^{2} \RP \gg 1 $. We note that $\Pcoupling^{\star}$ decreases with increasing refractory period. For $L=200$, we predict that the transition would occur at $\PCoupling^{\star} \approx 0.14 $ using $\PDefect=0.05$ and $ \RP=50$. The calculation of $P_{\text{risk}}$ can be extended to multiple cell layers (see supp. materials) and the abrupt transition remains.



Anti-arrhythmic drugs which prolong the refractory period increase the wavelength of reentrant ${\theta}{^\prime} \times \RPprime$, thereby reducing the number of possible independent electrical wavefronts which can propagate in the atrium. This reduces the complexity of the fibrillation. However, in addition to this, our model also predicts that prolonging the refractory period $\RP \!\to\! \tilde{\RP} > \RP$ reduces the structure-related risk of inducing fibrillation by rendering once arrhythmic regions, $\Pcoupling < \Pcoupling^{\star}(\RP)$, non-arrhythmic, $\Pcoupling > \Pcoupling^{\star}(\tilde{\RP})$. This implies that functional changes can render structurally arrhythmic regions non-arrhythmic without actually modifying the structure. 

The incidence of AF increases with age and is associated with the accumulation of fibrosis. Our objective was to gain insight into the mechanisms responsible for the initiation and maintenance of AF in this context using a phenomenological model. This deliberately simple model incorporates pertinent basic properties of tissue architecture and electrical dysfunction, excluding electrophysiological properties normally thought to be important in fibrillation such as the dispersion of refractoriness and restitution. The most interesting aspects of this study are: (1) The mere simplicity of the model. (2) The integration of simplified dynamics on a relevant discrete medium mimicking the anisotropic branching architecture of heart muscle. (3) A single model that reproduces multiple well known observed features of AF. (4) The spontaneous emergence of AF (without external stimuli) when the transversal coupling is reduced, mimicking fibrosis, beyond a threshold value. (5) Identification of local critical regions initiating and maintaining AF. (6) Analytical tractable risk of AF. (7) The existence of an analytically tractable threshold value of heterogeneity below which AF emerges, providing insight into how to control the risk. (8)  A targeted ablation procedure is possible for those patients at the early onset of AF driven by fibrosis due to the small number of critical regions.

It would be of interest to investigate a non-uniform distribution of vertical coupling (e.g., localised fibrosis) or refractory period \cite{krinskii1966spread}. The present model identifies specific targets for ablation based on the underlying structure of the myocardium. Patterned cultures of cardiac myocyte monolayers or cardiac tissue slices could test the primary predictions of the model as stated above. If the biological experiments agree with the model predictions then such targets within a biological system might be characterised electrophysiologically or architecturally with the aim of translating such insight into clinical applications thereby, hopefully, improving the success rate of treating AF with ablation.

\begin{acknowledgments}
\textit{Acknowledgements}: We thank Zuzanna Zukowska for helping implement the model. This work was supported by the British Heart Foundation. \textit{Author Contributions}: K.C. conceived the model. K.A.M. and K.C. performed the analytic calculations. K.A.M. implemented the model, collected and analysed the data and produced the figures. N.S.P. provided advice on data analysis. All authors contributed to writing the manuscript.
\end{acknowledgments}

\bibliographystyle{apsrev}
\bibliography{References}

\end{document}